\documentclass[11pt]{article}
\usepackage{amsmath}

\title{Superstring representation of Hubbard model}
\author{\large V. M. Zharkov \\[3mm]
\em Laboratory of Organic Semiconductors,  \\
\em Institute of Natural Sciences of Perm University, \\
\em Genkel st. 4, Perm, 614600, Russia}

\date{}
\begin{document}
\maketitle
\begin{abstract}

The new method of investigation of strongly correlated electronic system is
presented. Using this method we have derived the superstring from 2D Hubbard
model. The novel expression for generalized supercoherent state has been
calculated.

\end{abstract}

{\bf 1.} Nowadays the Hubbard model attracts considerable attention. This
model is incorporated in several profound theoretical projects, proposed in
\cite{anderson-91} for example. These projects lead to deep interaction of
ideas and methods of modern field theory and solid state physics \cite
{fradkin-91}. In this article we present the new approach to the strongly
correlating electronic systems. Using the functional integral we have
derived the superstring from 2-D Hubbard model. Essential ingredient of this
approach is the using of the supercoherent state, which exact expression was
calculated.

{\bf 2.} Let us recall the formalism developed in \cite{zharkov-1} for
systems with coulomb interactions.The system under consideration is
described by Hubbard model:

\begin{equation}
H=-t\sum_{<{\bf r},{\bf r^{\prime }>,s }}\,\alpha {}_{{\bf r,s }%
}{}^{+}\,\alpha {}_{{\bf r^{\prime },s }}{}\,+\,U\sum_{{\bf r}}\,n{}_{%
{\bf r},\uparrow }n{}_{{\bf r},\downarrow }=
\end{equation}
\begin{equation}
\label{hubb}\sum_{{\bf r}}U\,X{}_{{\bf r}}{}^{22}+\sum_{A,C,{\bf r},{\bf %
r^{\prime }}}\,t{}_{-A}{}_C({\bf r}-{\bf r^{\prime }})\,X{}_{{\bf r}%
}{}^{-A}\,X{}_{{\bf r}^{\prime }}{}^C
\end{equation}
where $<r,r^{\prime }>$-denote the sum over the nearest neighbors. Electrons
are described by $(\alpha _{rs }^{+},\alpha _{rs })$ and live in
two dimensional plane $r=\{x,y\}$ (for example square lattice), s  -
spin of the electrons. We give two different forms of this model (the first
form is a standard, the second contains the Hubbard operators $X^A$ (in fact
they are projectors $X_r^A=\mid pr><qr\mid $). These operators act in space
of following states: $\mid \,0\,>,\mid \,\uparrow \,>=\alpha _{\uparrow
}^{+}\,\mid \,0\,>,\mid \downarrow \,>=\alpha _{\downarrow }^{+}\,\mid
\,0\,>,\mid \,2\,>=\alpha _{\uparrow }^{+}\,\alpha _{\downarrow }^{+}\,\mid
\,0\,>$ (the space of eigenfunctions of Hubbard repulsion-the main and most
complicated interaction in many interesting models), $A=(p,q);p,q=0,\uparrow
,\downarrow ,2$. $X^{pq}$ has one nonzero element, sitting in p row and q
coloum of $4\times 4$ matrix. Local state of system and the Hilbert space
 we describe by supercoherent state:$\mid G[\chi (r,t^{\prime
}),E(r,t^{\prime }),h(r,t^{\prime })>$ where $\chi =\{\chi _1,\chi _2\}$ is
dynamical two component odd valued grassmanian fields,$E=\{E_1,E_2,E_3\}$
and $h=\{h_1,h_2,h_3\}$ are three component even valued dynamical fields. $%
\chi $ describes electronic degree of fredoms, $E$ and $h$ parametrises
charge density and spin density fluctuations appropreatly. According to this
formulation the many particle systems are described by following effective
functional:
\begin{equation}
\label{act}L_{eff}={\frac{{<\,G(\theta ,{\bf r},t^{\prime })\,\mid }\,\left(
{\frac{{\partial }}{{\partial t^{\prime }}}}-H\right) \,\mid G(\theta ,{\bf r%
},t^{\prime })\,>}{{<\,G(\theta ,{\bf r},t^{\prime })\,\mid \,G(\theta ,{\bf %
r},t^{\prime })\,>}}}
\end{equation}
where $\mid \,G\,>$ -is supercoherent state, which is expressed through
generators of the dynamic superalgebra, $\{{\bf r},t^{\prime },\theta \}$
-supercoordinates of superspace. The classification of various types of
(super)algebra in interactive systems was given in \cite{zharkov-2}. H is
the Hubbard hamiltonian expressed through the infinite dimensional
superalgebra generators $\{\,X_{{\bf r}}^\alpha \,\}$:  where $U-$ on-site
coulomb repulsion, $t_{\alpha ,\beta }({\bf r-r^{\prime }})$ -
''interaction'', arisen from kinetic energy. $\mid \,G\,>$ can be
constructed by following way \cite{zharkov-1}:
\begin{equation}
\label{corr}\mid G\,>={e^{-\sum_{k=1}^6X{}^k\,b{}^k({\bf r},t,\theta
)-\sum_{j=1}^2X{}^{-j}\,\chi {}^j({\bf r},t,\theta )}}\mid \,0\,>=\left(
F\chi ,z(E)+B\chi ^2\right)
\end{equation}
here $\mid 0>={\otimes }_{r}\mid 0>_r$, $\{b{}^C\}=\{\{E_i\},\{h_i%
\}\}$ $i=1,2,3$ and $\{X^k\}=\{X^{02},X^{20},X^{00}-X^{22},X^{\uparrow
\downarrow },X^{\downarrow \uparrow },X^{\uparrow \uparrow }-X^{\downarrow
\downarrow }\},$

$\{X^1,X^2\}=\{X^{0\uparrow }+X^{\downarrow 2},X^{0\downarrow }-X^{\uparrow
2}\}$- set of the bosonic fields (even valued grassmanian fields).
We have following expansion for unity:$\int \mid
G><G\mid d\mu (O)=1$ where $\mu (O)=<G\mid G>d^2\chi d^3Ed^3h$ is measure of
all dynamical fields $O=\{O_s\}=\{\chi ,E,h\}$ $\mid G>$ has four component,
two of them are fermionic (odd valued grassmanian nonlinear composite
fields), and two are bosonic (also composite and nonlinear in $\chi ,E,h$).
 Expanding the
expression (\ref{corr}) in the form of infinite series and then summing some
definite series that are matrix element of (\ref{corr}) we obtain following
nonlinear expression:
\begin{equation}
\label{1}F=a^{\prime \prime }+a^{\prime }\,E_z+{\bf h}\,{\bf \sigma ^P}%
\,\left( a^{\prime }+a\,E_z\right) ,\quad B=\left( \{0,a\}+\rho (\psi
)\,z(\tau )\,E{}^{+}\right)
\end{equation}

where the prime means differentiation with respect to $\delta $,\thinspace $%
\sigma _\mu ^P$ -Pauli matrix.
$$
z(\tau )=\{\cos (\tau )+\cos (\theta ^{\prime })\,\sin (\tau ),{e^{i\,\phi
^{\prime }}}\,\sin (\tau )\,\sin (\theta ^{\prime })\},\tau ={\it \arcsin }({%
\ \frac{{E\,\psi }}\rho }),
$$
$$
\rho ={\sqrt{{E^2}\,{{\psi }^2}+{{\psi ^{\prime }}^2}},}\qquad \psi ={{%
\delta }^5}\,{\frac{{\partial \,f\,}}{{\partial \,E^2}}},a={\frac{{{{%
\partial }^2f}}}{{{E^2}\,{{\partial \delta }^2}}}}
$$
\begin{equation}
\label{2}\qquad f=-{h^{-2}}+{\frac{{{E^2}\,\left( {\frac{{\sin (E\delta )}}{{%
E^3}}}-{\frac{{\sin (h\delta )}}{{h^3}}}\right) }}{{{E^2}-{h^2}}}}
\end{equation}
We use the spherical coordinate system $\{\,E,\theta ^{\prime },\phi
^{\prime }\,\}$ for {\bf E}. In (\ref{1},\ref{2}) we put $\delta =1$ after
calculation.

{\bf 3.} The partition function of the system can be written as:
$$
Z_{hub}=\int D\chi {}^{*}\,D\chi \,D{\bf E}\,D{\bf h}\,{e^{-i \int
d{}^2r\,dt^{\prime }\,d^2\theta \,{e^{\theta \,\theta {}^{*}}}\,L({\bf E},%
{\bf h},\chi )}} $$ As usual \cite{fradkin-91},\cite{zharkov-1}
,we divide the time interval into N equal parts and finally take
the limit:$N\rightarrow \,\infty $. Turning to three dimensional
lattice we obtain following representation for:

\begin{equation}
\label{pro}{e^{\int -i d^2{\bf r}\,dt^{\prime }\,L}}=\prod_{n,m,k}{%
e^{-i\,L(n,m,k)\Delta }}=\prod_{n,m,k}\left( 1-i\,L(n,m,k)\Delta \right)
\end{equation}
\thinspace where \thinspace $\{\,{\bf r},t^{\prime
}\,\}=\{x,y,t^{\prime }\}=\Delta \,\{n,m,k\,\}$. Now our intent is
to show that some functional variables (for example
$\{E{}^{+},E{}^{-}\}$ can be exchanged by variables in space-time
integral $(x,y,t^{\prime })$). As a result of this transformation
we will get some new functional representation for the partition
function with different relations of the functional and the
ordinary coordinates of integration. This representation will be
the partition function of a superstring. Let us take the measure
of fields component and represent this expression in following
manner: $\int DE{}^{+}=\prod_x\int dE{}_x^{+}$. Combining it with
the expression (\ref {pro}) and expanding the infinite product to
sum we can obtain the expression in first order of derivative:

\begin{multline}
\int \prod_x\left( 1-iL(x)\Delta \right) \,dE{}^{+}=\sum_x\Delta
\,\left( 1-i\int dE{}^{+}\,L(E^{+}(x))\right) =\\
\int DX\,{e^{-i\int dE^{+}\,L(E^{+})}}
\end{multline}

Now we consider all fields $\{\chi ,{\bf h},E_z\}$ as functions of $%
\{E^{+},E^{-}\}$ only, for example: $E_z=E_z(\theta ^{^{\prime }},\phi
^{\prime })$. In such way we get the reduction of the space-time variables.

{\bf 4.}Let us construct the map of two dimensional sphere parameterized by $%
\{E{}^{+},E{}^{-}\}$ (we can as well select $\{h^{+},h^{-}\}$, instead of $%
E^{\pm }$ ) to three dimensional space $R^3$. Now we exploit the results
deduced in \cite{budinich-86},\cite{vis-91} for the conformal immersion
of two-dimensional surface that is string world sheet in $R^3$. We use the
Gauss map in the same manner, as in \cite{budinich-86}. This map is defined
in our case by following procedure:

i) one make takes the two-dimensional sphere to be parameterized by spinor $%
z:(zz{}^{*}=1)$. This Riemann sphere gives conformal compactification of
two-dimensional plane and because of this it describes the superstring world
sheet. Having the spinor $z$ we can construct complex null three-dimensional
vector:
\begin{equation}
\label{t}t{}_\mu ^{+}{}=\overline{z}\,\sigma {}_{_\mu }z,\qquad t_\mu
^{-}=(t_\mu ^{+})^{\bullet }{}
\end{equation}
The important feature of null vector for us is that:
\begin{equation}
\label{null}{{t{}_1}^2}+{{t{}_2}^2}+{{t{}_3}^2}=0
\end{equation}

ii) For string, propagating in three-dimensional space, it is known that the
ivolution equation takes the form:
\begin{equation}
\label{string}{\frac{{{\partial }^2\,X{}_\mu }}{{\partial \,\sigma
^{+}\,\partial \,\sigma ^{-}}}}=0\qquad \left( \partial _{\pm }X^\mu \right)
^2=0
\end{equation}
It is obvious from (\ref{string}) that real vectors ${\frac{{\partial
\,X{}_\mu }}{{{\partial }\,\sigma {}^{\pm }}}}$ are null.

iii) From (\ref{t}) we see that vector ${\frac{{\partial \,X{}_\mu }}{{{%
\partial }\,\sigma {}^{\pm }}}}$ and $t_\mu ^{\pm }$ are tangent vectors to
string world sheet. This similarities between (\ref{null}) and (\ref{string}%
) gives us the possibility to construct the mapping of ${R^2}$ to
${R^3}$ (Gauss map):

$$
\partial _{\pm }\,X^\mu =\Omega \,t_\mu ^{\pm }
$$ Having obtained the Gauss map the question of reconstructing of
the functional variables (coordinates of string world sheet) is
answered by expression: $X{}_\mu (\sigma {}^{+},\sigma
{}^{-})={\rm Re}(\int^\sigma
d\zeta \,\Omega (\zeta )\,t{}_\mu )+X{}_\mu (0,0)$ \cite{budinich-86} ,%
\cite{vis-91} where $\Omega $ is given by equation: $$
{\it \ln }(\Omega ){}_\sigma =-{\frac{{t{}^\mu \,t{}_\sigma {}^\mu }}{\mid {%
t\mid ^2}}}
$$
Here the $\sigma $ index means the derivative on $\sigma $. This equation is
the integrability condition of the Gauss map.

{\bf 5.}Inserting $t{}_\mu {}^{{\pm }}$ into the following
lagrangian: $$ 2\int d^2\sigma \,t{}_\mu {}^{-}\,t{}^{+}{}_\mu
=\int d^2\sigma \sqrt{\left( t{}{}^{+}-t{}{}^{-}\right) ^2\left(
t{}{}^{+}+t{}{}^{-}\right) ^2}\,\,= $$
\begin{equation}
\label{NG}\int d^2\sigma \,{\sqrt{{{\stackrel{.}{X}}^2}\,X^{\prime ^2}-({{%
\stackrel{.}{X}X^{\prime }{})}^2}}}
\end{equation}
we see that (\ref{NG}) is a Namby-Goto string \cite{budinich-86}. For our
purpose it is convenient to take this lagrangian in form given by Polyakov
\cite{polyakov-1}:
\begin{equation}
\label{pol}L{}_{N-G}=\,{{\sqrt{g}}\,g{}^\alpha {}^\beta }\,{\frac{{\partial
\,X{}_\mu \,\partial X{}_\mu }}{{{\partial }\,\sigma {}^\alpha \,\partial
\,\sigma {}^\beta }}}
\end{equation}
where $g_{\alpha \beta }$ metric tensor of surface . Taking the $\widetilde{%
g{}}^\alpha {}^\beta ={\frac{{\partial \,X{}_\mu \,\partial \,X{}_\mu }}{{{%
\partial }\,\sigma {}^\alpha \,\partial \,\sigma {}^\beta }}}$ we find that $%
g_{\alpha ,\beta }$ can be expressed in following manner:
$g{}_{\alpha ,\beta }={\rm C}\,\widetilde{g{}}_{\alpha ,\beta }$
where C is arbitrary constant. It is seen that we deal with the
conformal gauge for $\widetilde{g} $:
$\widetilde{g}_{12}={\frac{-(z_1z_2^{*}\,+\,z_2z_1^{*})^2}{{2}}},
\widetilde{g{}}_{11}=\widetilde{g{}}_{22}=0$. Now let us show how
to derive
the bosonic sector of superstring. Evaluating expression ($z_{{\bf r}%
}\,\left( \sum_{{\bf r}}U\,X{}^{{\it pp}}{}_{{\bf r}}\right) \,z{}_{{\bf r}%
}^{*}$ ) we get:
\begin{equation}
\label{lo}L{}_0=\kappa _1\left( 1-2\,\sin (e)\,E{}^{+}E{}^{-}\right)
\,\left( 1+2\,\left( a^{\prime \prime }-{{a^{\prime }}^2}\,E{{_z}^2}\right)
\right)
\end{equation}
Chemical potential $\epsilon $ gives the additional term: $\kappa _2$$npn$ .
Substituting expression of $t_\mu ^{\pm }$ to (\ref{NG}) we obtain any
function of $E_z(\theta ^{\prime },\phi )$:

By adding the lagrangian (\ref{pol}) to (\ref{NG}) we can obtain the string
lagrangian if (\ref{lo})is constrained by equation: $L_{N-G}\,+\,\sqrt{g}%
\,+\,L_0=0$

Returning to (\ref{hubb}) we see that the Hubbard repulsion in continuum
limit can be transformed to the string lagrangian:
\begin{equation}
\label{Lsite}L_{site}=\,{\sqrt{g}}+L{}_{N-G}
\end{equation}

{\bf 6.}\thinspace Effective
functional can be evaluated as a series of a grassmanian fields:
\begin{equation}
\label{Leff}L_{{\sl eff}}=L_0+L_2+L_4
\end{equation}
Combining kinetic energy and (\ref{Lsite})  the following result can be
obtained:

$
\begin{array}{c}
L=\,
{\sqrt{g}}\,g{}^\alpha {}^\beta \,{\partial }{}_\alpha X{}^\mu \,{\partial }%
{}_\beta X{}^\mu +\,{\sqrt{g}}+\,\Upsilon {}^\alpha {}^\beta {}_i{}_j\,{%
\partial }{}_\alpha \phi {}^i\,{\partial }{}_\beta \phi {}^j+ \\ \Psi
_e^{\bullet }(e^\alpha \partial _\alpha +\Pi )\Psi _e+R_{ijkl}\chi
_i^{*}\chi _j^{*}\chi _k\chi _l
\end{array}
$

where $\Psi _e=\{\chi ,\chi ^{\bullet }\}$ is two component spinor, and $%
\gamma _i$ act in space of this components, $\,\Upsilon {}^\alpha {}^\beta
{}_i{}_j\,{\partial }{}_\alpha \phi {}^i\,{\partial }{}_\beta \phi
{}^j=(1-npn)\nabla ^2A+\nabla ^2npn$ and $e^\alpha ,\Pi ,R_{ijkl}-$are given
in first order of coefficients of superfield ,(the full
expression for them are complex and will be given in other paper).

$$
\begin{array}{ccc}
e^\alpha & =(e)_1+(e)_2= & [C_{12}^{12}(1-\sigma ^1)\gamma
^5-(C_o^{12}\gamma ^1+c.c.)]+n\gamma ^5... \\
\Pi & = & (t\nabla -\kappa _1)(e)_1+(t\nabla -\kappa _2)(e)_2+... \\
R_{1221} & = & (t\nabla -\kappa _1)C_{12}^{12}-(t\nabla -\kappa _2)\det
n+...
\end{array}
$$

where $C_{12}^0=2t(z_1B_2)^{*}det(F)\,+\,\kappa _1\,z^{*}\gamma
^5B,\,C_{12}^{12}=2t(det(F))^{*}z\gamma ^1B\,+\,\kappa _1B^{*}\gamma
^5B\,+\,c.c,\,n_{12}^0=z^{*}B,\,n_{12}^{12}=B^{*}B,\,n=F^{+}F,\,npn=%
\sum_{i,j=1}^2n_{ij,}A=z^{*}\gamma ^5z,\kappa _1=U/2-\epsilon ,\kappa
_2=2\epsilon -U/2$.

$\,$ The operator $\nabla $ equal to $\frac \Delta \Omega \sum_{i=\pm
}\left[ \sum_{\mu =1}^3\left( \frac{\partial X_\mu }{\partial \sigma _i}%
\right) ^{-1}\right] \frac \partial {\partial \sigma _i}$ and acts on the
right superfunction in any terms of the effective action: $\nabla (\Phi
^{*}\Phi )=\Phi ^{*}\nabla \Phi $. This condition follows from correct
determination of continuum limit of kinetic energy.

\end{document}